\documentclass{PoS}
\usepackage{amsmath}
\title{AGN observations with a less than 100 GeV threshold using H.E.S.S. II}

\ShortTitle{AGN observations with a less than 100 GeV threshold using H.E.S.S. II}

\author{\speaker{Dmitry Zaborov}\thanks{On leave from Institute for Theoretical and Experimental Physics, B. Cheremushkinskaya 25, Moscow, 117218 Russia}\\
Laboratoire Leprince-Ringuet, \'Ecole Polytechnique, CNRS/IN2P3, 91128 Palaiseau, France\\
E-mail: \email{zaborov@llr.in2p3.fr}}

\author{Carlo Romoli, Andrew M. Taylor\\
Dublin Institute for Advanced Studies, 31 Fitzwilliam Place, Dublin 2, Ireland}

\author{Jean-Philippe Lenain\\
LPNHE, Universit\'e Pierre et Marie Curie Paris 6, Universit\'e Denis Diderot Paris 7, CNRS/IN2P3, 4 place Jussieu, 75252, Paris Cedex 5, France}

\author{David Sanchez\\
Laboratoire d'Annecy-le-Vieux de Physique des Particules, Universit\'e Savoie Mont-Blanc, CNRS/IN2P3, F-74941 Annecy-le-Vieux, France}

\author{Robert Parsons\\
Max-Planck-Institut f\"ur Kernphysik, PO Box 103980, 69029 Heidelberg, Germany}

\author{for the H.E.S.S. collaboration}

\abstract{
The recent addition of the 28 m Cherenkov telescope (CT5) to the H.E.S.S. array extended the experiment's sensitivity towards low energies.
The lowest energy threshold is obtained using monoscopic observations with CT5, providing access to gamma-ray energies below 100~GeV.
This is particularly beneficial for studies of Active Galactic Nuclei (AGN) with soft spectra and located at redshifts $\geqslant$ 0.5.
Stereoscopic measurements with the full array (CT1-5) provide a better background rejection than CT5 Mono, at a cost of a higher threshold.
We report on the analysis employing the CT5 data for AGN observations with a < 100~GeV threshold.
In particular, the spectra of PKS 2155--304 and PG 1553+113 are presented.}

\FullConference{The 34th International Cosmic Ray Conference \\
		 30 July- 6 August, 2015\\
		 The Hague, The Netherlands}

\begin{document}

\section{Introduction}
The High Energy Stereoscopic System (H.E.S.S.) is a system of Imaging Atmospheric Cherenkov Telescopes located in the Khomas Highland of Namibia, 1800 m above sea level \cite{hess_crab_2006}.
The telescopes observe very high energy (VHE, E $\gtrsim$ 100~GeV) gamma rays via Cherenkov light emitted by the electromagnetic showers arising from gamma-ray interactions in Earth's atmosphere.
The images of the gamma-ray showers are recorded by cameras consisting of photomultiplier tubes (PMTs), which are installed in each telescope's focal plane.
During the first phase of the experiment (H.E.S.S. phase I), which lasted from 2002 to 2012, the system consisted of four 12 m telescopes, CT1-4.
A fifth telescope, CT5, of 28 m diameter, was added in 2012.
The larger reflector area of CT5 (596 m$^2$) as compared to CT1-4 (107 m$^2$) results in a lower energy threshold.
The imaging of low energy showers is also improved by using a more finely pixelated camera (960 PMTs for CT1-4, 2048 PMTs for CT5).
The five-telescope configuration is known as H.E.S.S. phase~II, or H.E.S.S.~II.

The trigger system of H.E.S.S.~II supports two types of physics triggers - Mono and Stereo.
The Mono trigger uses exclusively the information from CT5.
The Stereo trigger requires at least two telescopes out of five.
Because the field of view of CT5 is smaller than for CT1-4 (1.5$^\circ$ radius instead of 2.5$^\circ$), the stereo triggers do not always include CT5.
The standard observation mode of H.E.S.S.~II is to collect both Mono and Stereo events during the same data taking run.

The analysis of CT1-5 Stereo data provides a somewhat lower energy threshold, better hadron rejection and better angular resolution than with CT1-4 only.
The energy threshold of the CT1-5 Stereo analysis is still largely limited by the optical sensitivity of CT1-4.
The analysis of CT5 Mono events provides a much lower energy threshold than CT1-5 Stereo.
However the absence of stereoscopic constraints makes the rejection of hadronic events more difficult, leading to a larger background and reduced signal-to-background ratio at the analysis level.
The low energy threshold of CT5 Mono implies high event rates, thus small statistical errors, which leads to tight requirements for the accuracy of background subtraction.
The angular reconstruction of the monoscopic analysis is significantly less precise than in Stereo mode.
For point-like sources this leads to a reduction in sensitivity.
Nonetheless, the CT5 Mono analysis provides unique opportunities for gamma-ray astronomy at energies $<100$~GeV, which are
complementary to satellite experiments ({\it Fermi}-LAT).
The low energy threshold provided by CT5 is particularly beneficial for studies of gamma-ray bursts
and active galactic nuclei (AGN) located at high redshifts ($z \gtrsim 0.5$),
and the spectral features caused by the gamma-ray interactions with the extragalactic background light.

This paper reports on the analysis of two bright AGN, PKS 2155--304 and PG 1553+113, using H.E.S.S.~II.
As compared to H.E.S.S. I observations
\cite{hess_pks2155_2005,hess_pks2155_2010_variability,hess_pks2155_long_term_monitoring,hess_pg1553_evidence,hess_pg1553_spectroscopy,hess_pg1553_flare},
the spectral measurements for both sources are extended towards lower energies.
The data sets and run quality selection are described in Sect.~\ref{sect:datasets}.
Section~\ref{sect:analysis} gives details on the data analysis technique.
The results for PKS~2155--304 and PG 1553+113 are reported respectively in Sect.~\ref{sect:pks2155} and~\ref{sect:pg1553}.
A cross-check analysis is briefly described in Sect.~\ref{sect:crosscheck} and conclusion are summarized in Sect.~\ref{sect:conclusion}.

\section{Data sets and run quality selection}\label{sect:datasets}
PKS 2155--304 was monitored with H.E.S.S.~II regularly between April and November 2013, further observations were taken in May and June 2014.
PG 1553+113 was observed with H.E.S.S.~II between May 30, 2013 and Aug 7, 2013.
To ensure the quality of the data sets and their fitness for the CT5 Mono analysis the following run quality criteria were applied.
\begin{itemize}
\setlength\itemsep{0em}
\item Source position located between 0.35 and 1.2$^\circ$ off-axis from camera center;
\item Radiometer temperature less than 20$^\circ$ C and stable during the run within $\pm$ 3$^\circ$ C;
\item Relative humidity $<$ 90\%;
\item Run duration $>$ 5 min and live time fraction > 90\%;
\item At least 90\% of pixels in CT5 are functional;
\item CT5 trigger uses standard configuration, trigger rate between 1200 and 3000 Hz and stable within $\pm10$\%;
\item Telescope tracking functions normally.
\end{itemize}

\section{Data analysis}\label{sect:analysis}
The data sets were processed with the standard H.E.S.S. analysis software using the Model reconstruction \cite{model_reconstruction}
which was recently adapted to work with Monoscopic events \cite{model_reconstruction_hess_ii}.
The Model reconstruction performs a likelihood fit of the air shower image to
a semi-analytical model of an average gamma-ray shower parameterized as a function of energy, primary interaction depth, impact distance and direction.
Gamma-like candidate events are selected based on the value of the goodness-of-fit variable
and the reconstructed primary interaction depth.
In addition, events with estimated error in direction reconstruction $> 0.3^\circ$ are rejected.
The low energy threshold is controlled with a dedicated likelihood variable $NSBGoodness$.
Three standard cut configurations were defined, Loose, Standard and Safe, with different settings for the $NSBGoodness$ cut.
Loose cuts provide the lowest energy threshold, but may in some cases lead to elevated level of systematic errors in background subtraction.
Standard cuts provide a better control over the background subtraction at a cost of increased threshold.
Safe cuts are intended for performing cross-checks and analyzing complex regions with high level of night sky background.
The event selection cuts, except for the $NSBGoodness$ cut, were optimized for maximum significance on a point source with a photon index of 3.0
observed at zenith angle of 18$^\circ$.
The background subtraction is performed using the standard algorithms used in H.E.S.S.
-- the ring background method (for sky maps) and the reflected background method (with multiple OFF regions, for spectral measurements).
The ring background method was set up to use a zenith-dependent 2D acceptance model,
0.45$^\circ$ ring radius, 0.15$^\circ$ ring half-width, and oversampling radius of 0.1$^\circ$.
The angular distance cut for the reflected background method ($\theta^2$) was optimized for use with Safe cuts,
and re-used for Standard and Loose cuts.
This may lead to a $\sim$10\% loss of sensitivity but improves the signal-to-background ratio,
which is a key to reducing the effects of the background subtraction errors.
The CT5 Mono analysis was applied to all events that include CT5 data (ignoring information from CT1-4).


\section{PKS 2155-304}\label{sect:pks2155}
PKS 2155--304 is a high-frequency peaked BL Lac (HBL) object 
at $z = 0.116$ \cite{pks2155_redshift_Ganguly_2013}.
It was first discovered as a high energy emitter by the HEAO 1 X-ray satellite \cite{Griffiths_1979, Schwartz_1979}.
Gamma-ray emission in the energy range 30~MeV to 10~GeV was detected from PKS 2155--304 by the EGRET instrument aboard the Compton Gamma Ray Observatory satellite
\cite{Vestrand_1995}.
The first detection in the VHE range was attained in 1996 by the University of Durham Mark 6 Telescope \cite{Chadwick_1999a}. 
Starting from 2002 the source was regularly observed with H.E.S.S., with the first detection reported based on the 2002 data \cite{hess_pks2155_2005}.
Strong flux variability with multiple episodes of extreme flaring activity in the VHE band have been reported \cite{hess_pks2155_2010_variability, hess_pks2155_long_term_monitoring}.
A photon index of $3.53 \pm 0.06_{\text{stat}} \pm 0.10_{\text{syst}}$ was reported in \cite{hess_pks2155_2010_variability}
for the low flux state (2005--2007) above 200~GeV.
For average and high flux states the presence of a curvature or a cut-off was favored by the spectral fit \cite{hess_pks2155_2010_variability}.

\begin{figure}
  \centering
    \includegraphics[width=0.4\linewidth]{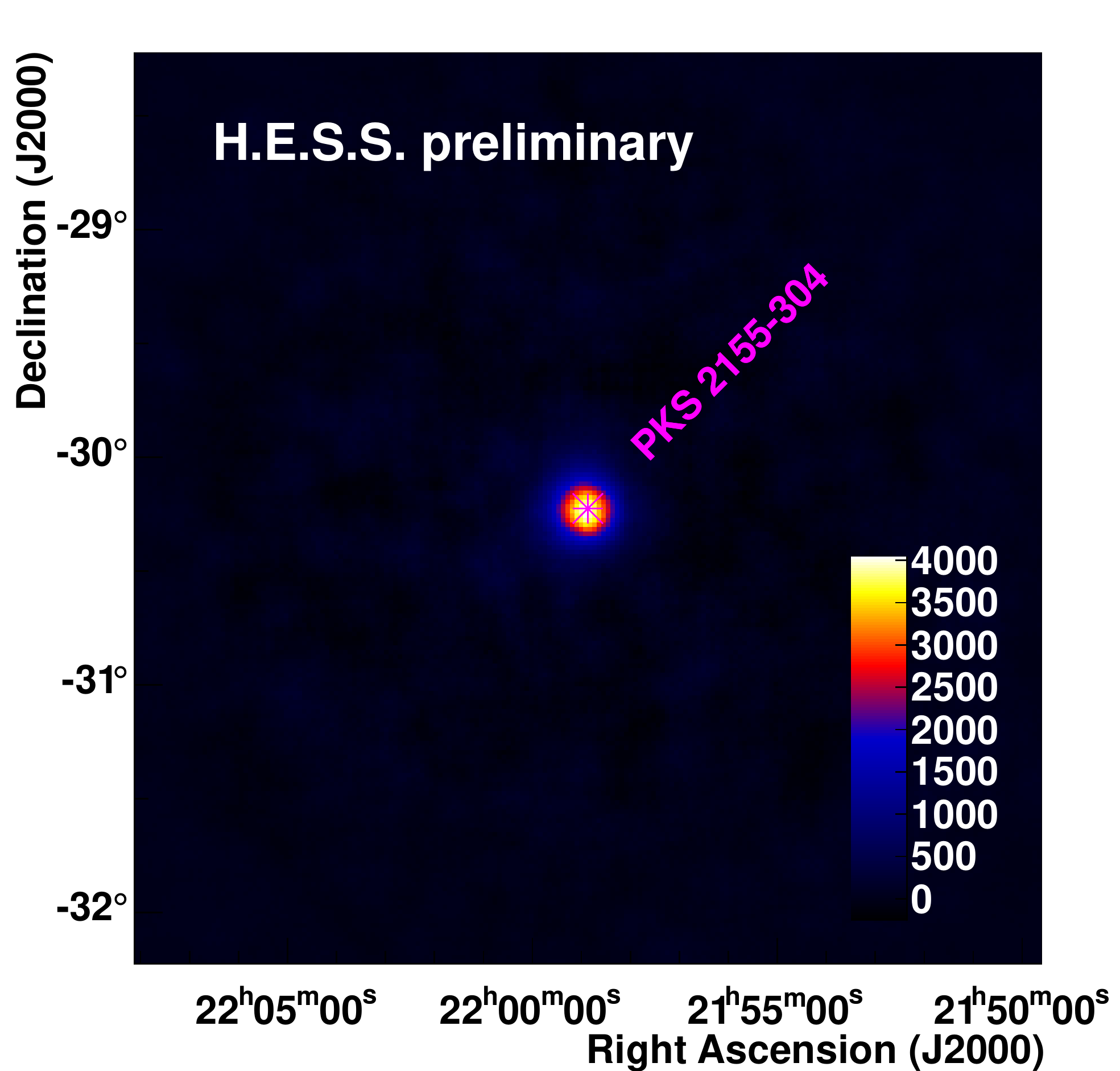}
    \includegraphics[width=0.4\linewidth]{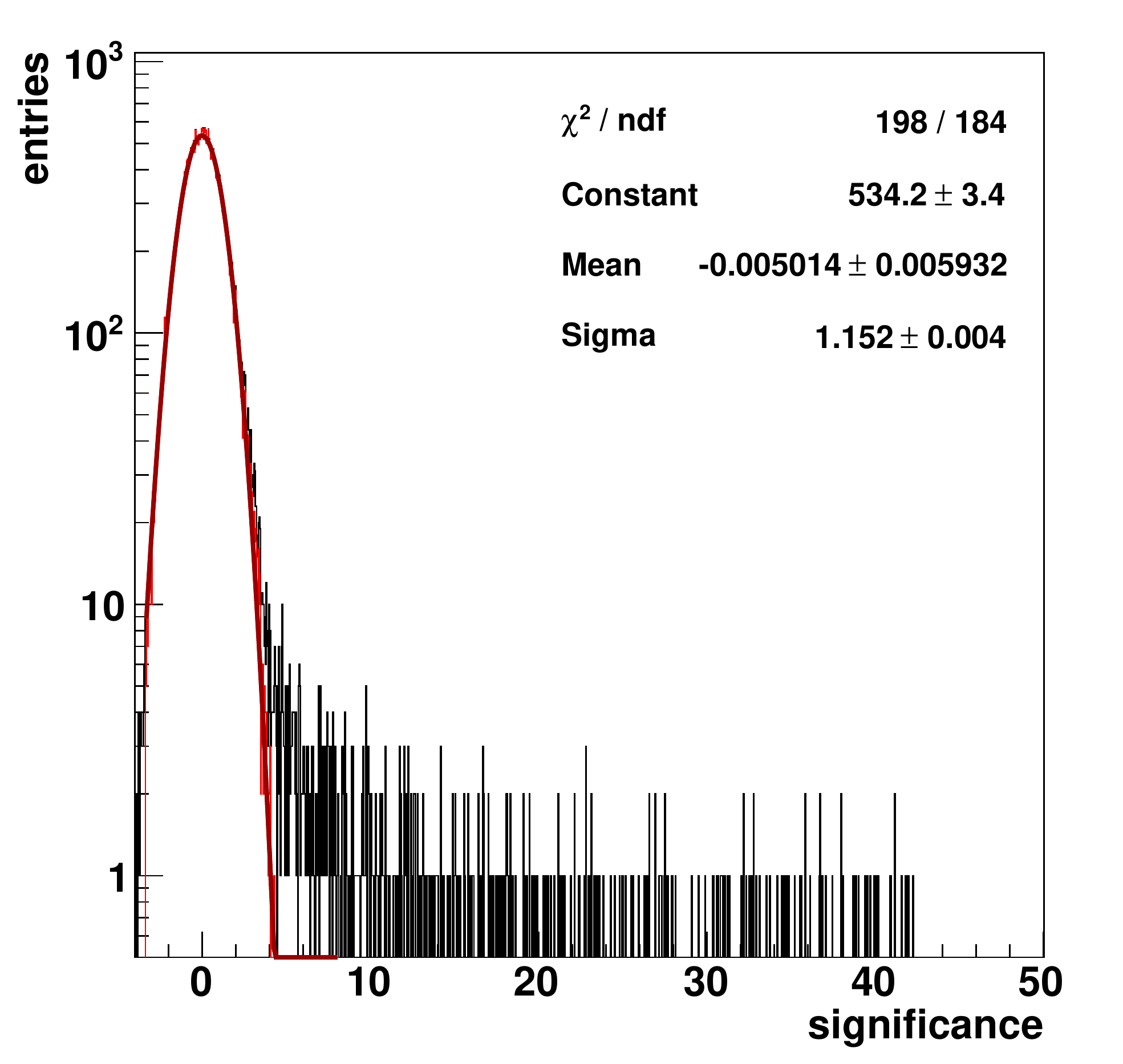}
  \caption{Left: A two dimensional distribution of excess events observed in the direction of PKS 2155--304
using the CT5 Mono analysis (2013--2014 data).
The source position is indicated by magenta star.
Right: The significance distribution that corresponds to the excess map (black histogram).
The distribution obtained by excluding a circular region of $0.3^\circ$ radius around the source is shown in red;
the results of a Gaussian fit to this distribution are also shown.
}
  \label{fig:pks2155}
\end{figure}

The PKS 2155--304 data set, filtered as explained in Sect.~\ref{sect:datasets}, comprises 138 data taking runs.
The total live time of the data set is 56.0 hr.
The observation zenith angle ranges from 7$^\circ$ to 60$^\circ$, with a median value of $16^\circ$.
This data set was analyzed using Standard cuts as described in Sect.~\ref{sect:analysis}.
The background event counts obtained for the OFF regions in each run (in the reflected background analysis) were used to perform an additional test of the uniformity of the camera acceptance.
This was done by computing, using a likelihood ratio test, the p-value of the hypothesis that the event counts observed in all OFF regions come from the same Poisson distribution.
The results of this test were found consistent with a uniform camera acceptance.
Only four runs had p-value < 1\%. 

\begin{figure}
  \centering
    \includegraphics[width=0.4\linewidth]{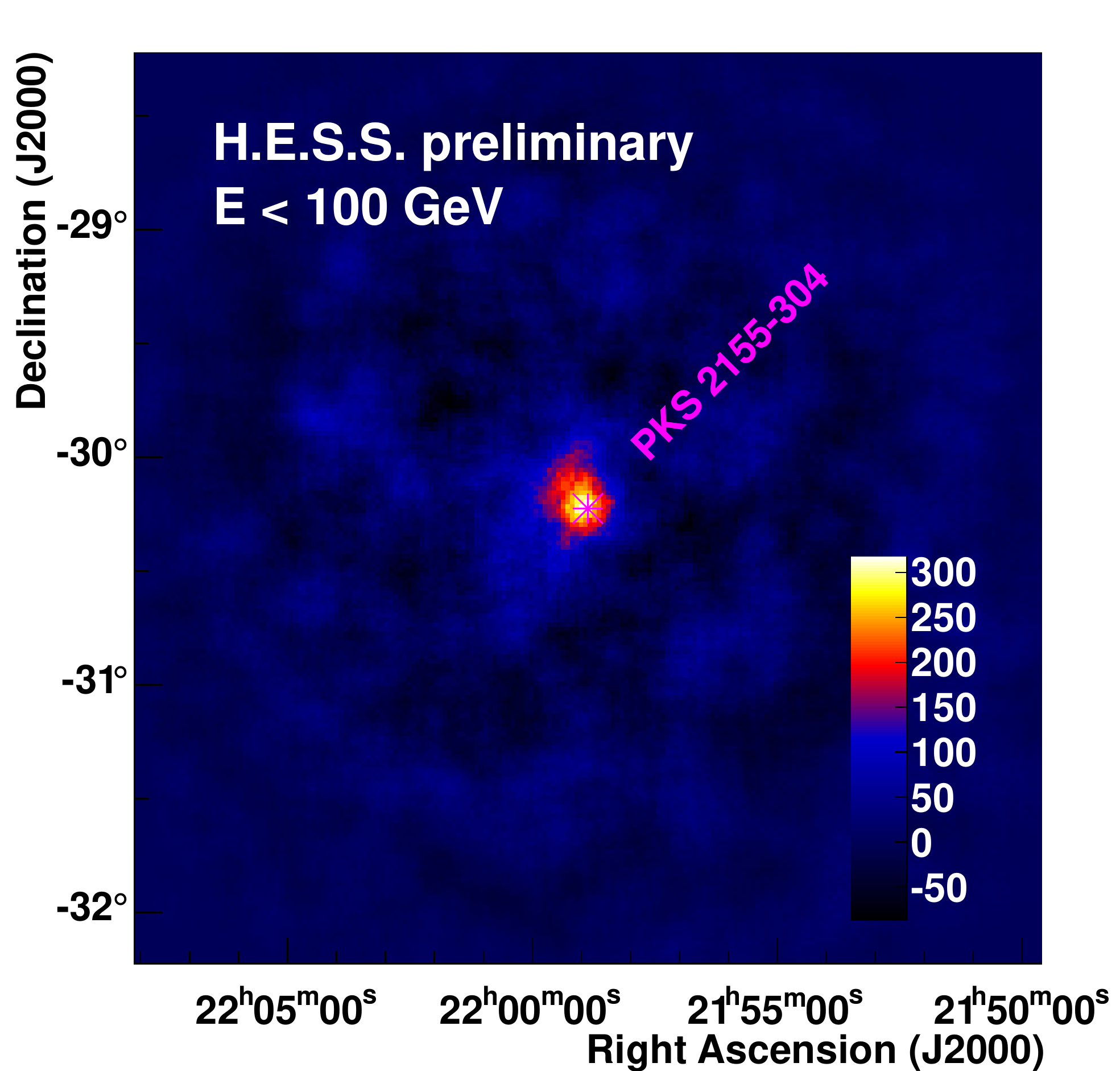}
    \includegraphics[width=0.4\linewidth]{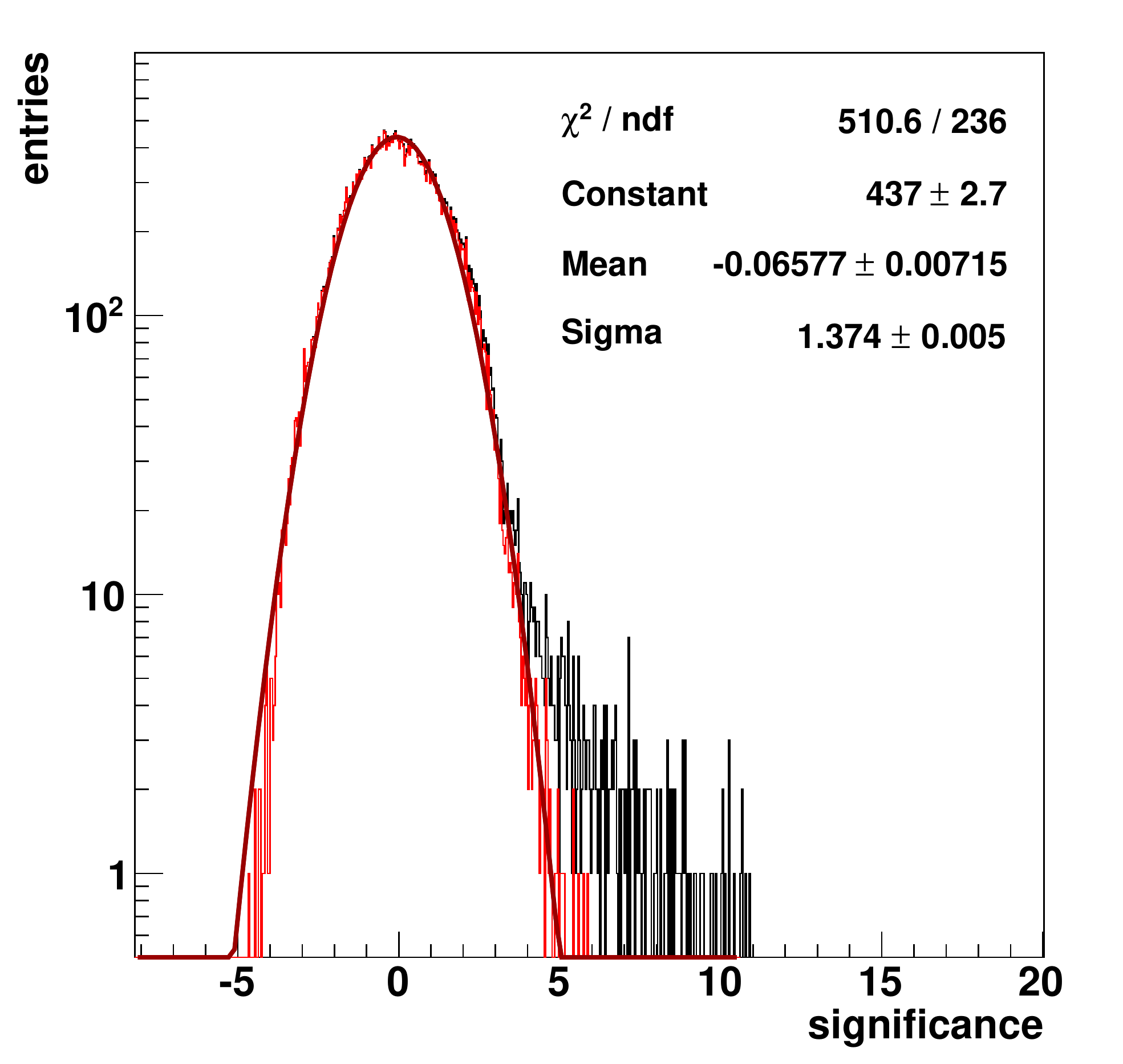}
  \caption{The PKS 2155--304 excess map (left) and significance distribution (right) for events with reconstructed energy E~<~100~GeV
(CT5 Mono analysis, 2013--2014 data).
}
  \label{fig:pks2155_ebins}
\end{figure}

The sky map obtained for PKS 2155--304 using the CT5 Mono analysis is shown in Fig.~\ref{fig:pks2155}, left.
The source is detected at $\approx$ 42 $\sigma$ significance, with at least 3500 excess events.
The corresponding distribution of the excess significance of all skymap bins is shown in Fig.~\ref{fig:pks2155}, right.
Outside the exclusion radius of $0.3^\circ$ the distribution is well fitted by a Gaussian with $\sigma = 1.15$.
This indicates the presence of a systematic effect in background subtraction,
whose $\sigma_{\text{syst}}$ corresponds to about 57\% of the statistical errors ($\sigma_{\text{stat}} = 1.0$).
This systematic effect is currently under study as part of a larger effort on understanding the mono analysis performance.
Repeating the analysis using only events with reconstructed energy below 100~GeV leads to a 10~$\sigma$ significance
at the position of PKS 2155--304 in the skymap (Fig.~\ref{fig:pks2155_ebins}).
The significance distribution outside the exclusion region has $\sigma = 1.37$.
Thus the source is confidently detected at $E < 100$ GeV.

\begin{figure}
  \centering
    \includegraphics[width=0.48\linewidth,height=0.31\linewidth]{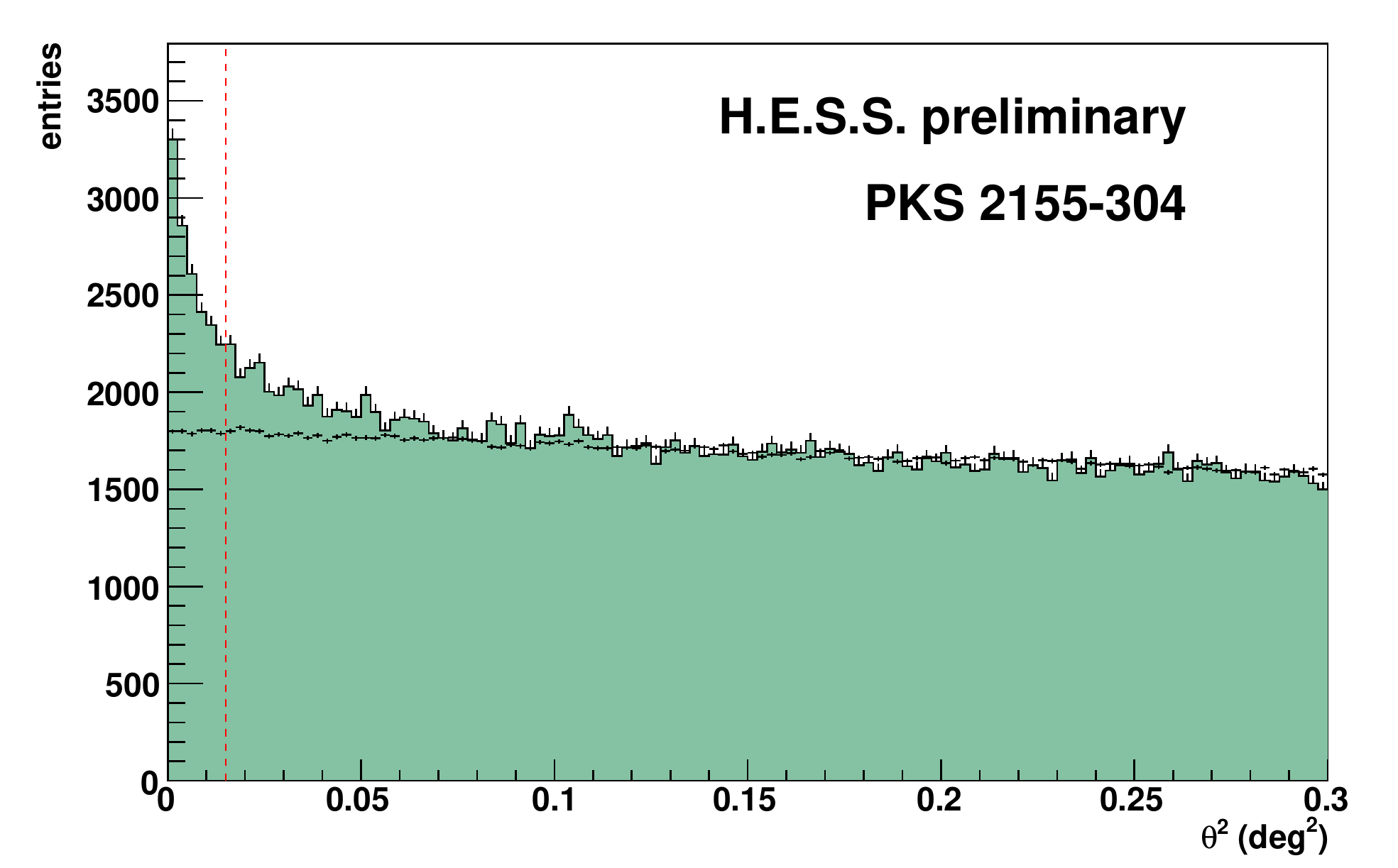}
    \includegraphics[width=0.50\linewidth]{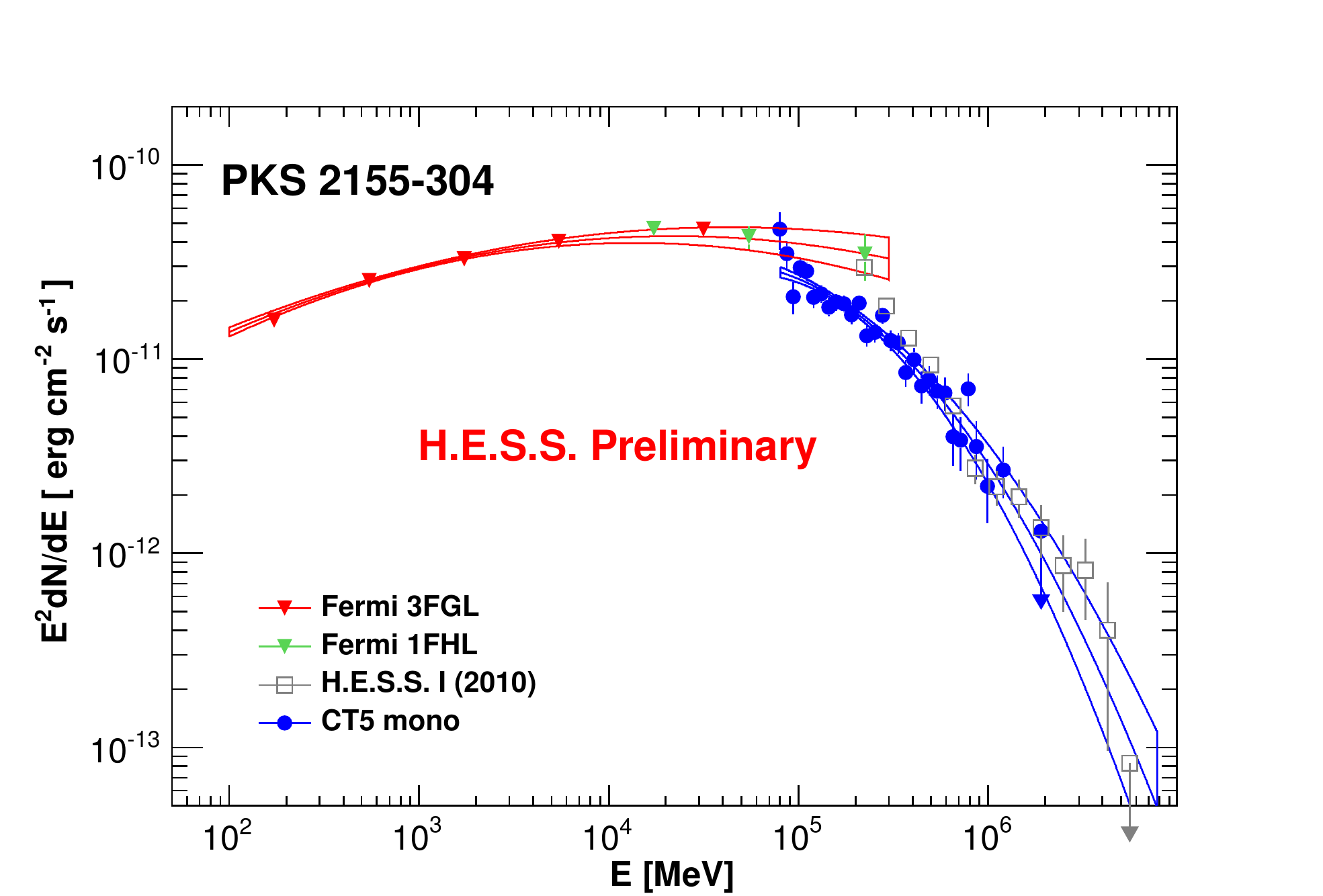}
  \caption{
Left: The distribution of $\theta^2$ (squared angular distance to PKS 2155--304) for gamma-like events obtained with the CT5 Mono analysis (filled histogram)
in comparison with the normalized $\theta^2$ distribution for off regions (black points).
The vertical dashed line shows the limit of the on-source region.
Right: The time-averaged energy spectrum of PKS 2155--304 obtained from the CT5 Mono analysis (2013-2014 data)
in comparison with the {\it Fermi} 3FGL and 1FHL catalog spectra \cite{3FGL,1FHL} and the archival H.E.S.S. data from \protect\cite{hess_pks2155_2010_variability}.
  }
  \label{fig:pks2155_theta2}
  \label{fig:pks2155_spectrum}
\end{figure}

The distribution of $\theta^2$, the square of the angular difference between the reconstructed shower position and the source position,
is shown in Fig.~\ref{fig:pks2155_theta2}, left (filled histogram).
A 42.9 $\sigma$ excess over the background (black crosses) is observed within the on-source region ($\theta^2 < 0.015$ deg$^2$).
The reconstructed spectrum of PKS 2155--304 is shown in Fig.~\ref{fig:pks2155_theta2}, right.
A log parabola\footnote{$dN/dE = \Phi_0\,(E/E_0)^{-a-b\cdot log(E/E_0)}$}
 is used to fit the data, with 
a flux normalization $\Phi_0 = (5.11 \pm 0.15_{\text{stat}}) \times 10^{-10} \, \text{cm}^{-2} \, \text{s}^{-1} \, \text{TeV}^{-1}$
at decorrelation energy $E_{0} = 156$ GeV,
photon index $a = 2.63 \pm 0.07_{\text{stat}}$ and curvature parameter $\beta = 0.24 \pm 0.06_{\text{stat}}$.
Systematic uncertainties of the spectrum parameters are under investigation and will be reported elsewhere
The spectrum data points (blue filled circles) cover the energy range from 80~GeV to 1.2~TeV (not including upper limits).
At $E > 300$~GeV the new measurement approximately matches the spectrum reported in \cite{hess_pks2155_2010_variability} 
for the quiescent state observed by H.E.S.S. in 2005--2007.
At $E < 300$~GeV the new spectral fit lies below the {\it Fermi}-LAT spectra reported in the 3FGL and 1FHL catalogs \cite{3FGL,1FHL}.
This is consistent with PKS 2155--304 being in a low flux state during the observations analyzed in this work.

\section{PG 1553+113}\label{sect:pg1553}
The HBL object PG 1553+113 was first established as a VHE emitter in 2005 by H.E.S.S. \cite{hess_pg1553_evidence}.
The H.E.S.S.~I measurements \cite{hess_pg1553_spectroscopy} yield a photon index $\Gamma = 4.5 \pm 0.3_{\text{stat}} \pm 0.1_{\text{syst}}$ above 225~GeV.
At high energies (HE, 100~MeV~<~E~<~300~GeV) the source was detected by {\it Fermi}-LAT
with a photon index of $1.68 \pm 0.03$ \cite{Fermi_bright_source_list,Fermi_pg1553}.
The redshift of PG 1553+113 is constrained by UV observations to the range $0.43 < z \lesssim 0.58$ \cite{pg1553_redshift_Danforth}.
Assuming that the difference in spectral indices between the HE and VHE regimes is due to the attenuation by the extragalactic background light,
the redshift was constrained to $z = 0.49 \pm 0.04$~\cite{hess_pg1553_flare}.

\begin{figure}
  \centering
    \includegraphics[width=0.4\linewidth]{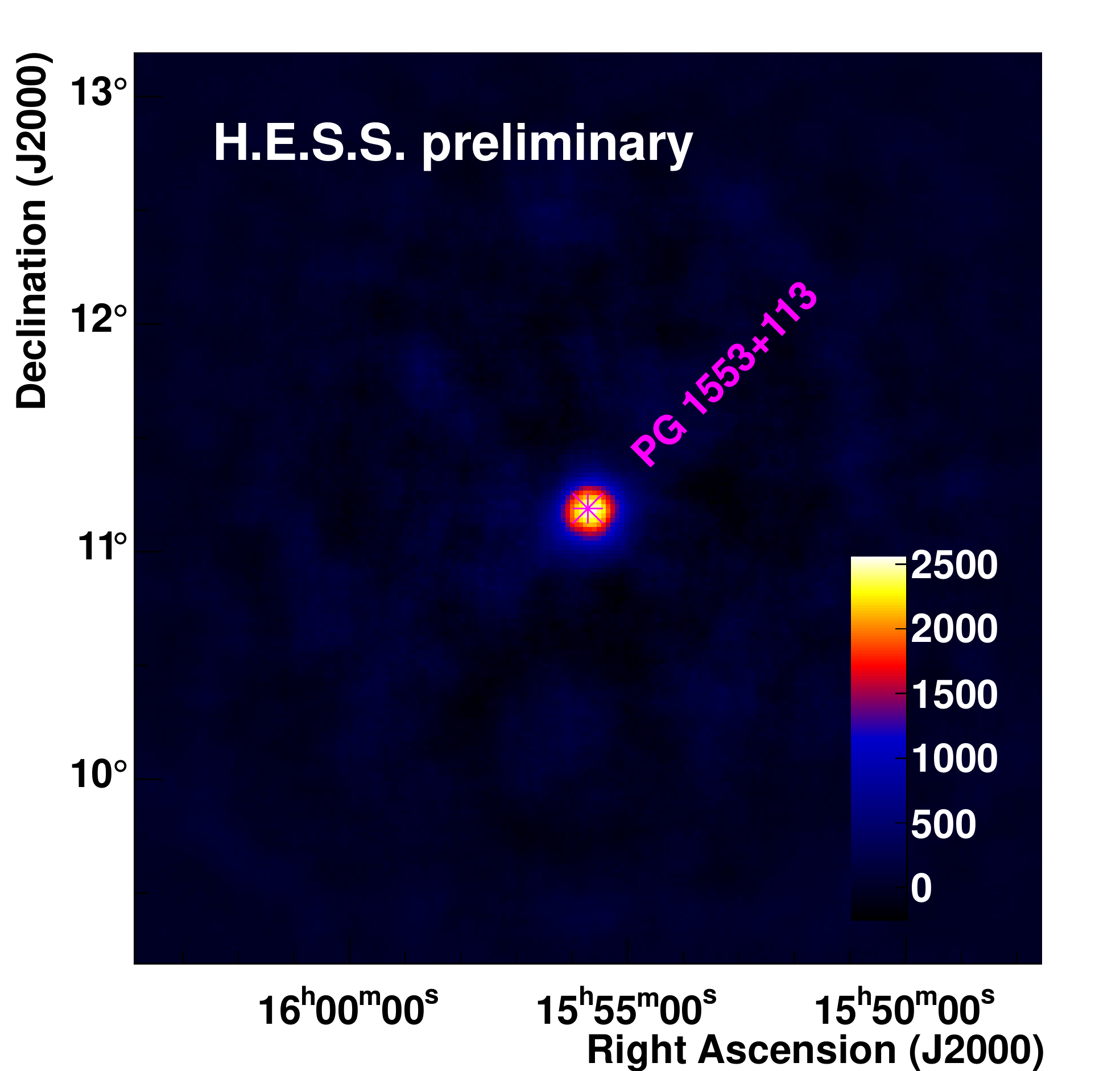}
    \includegraphics[width=0.4\linewidth]{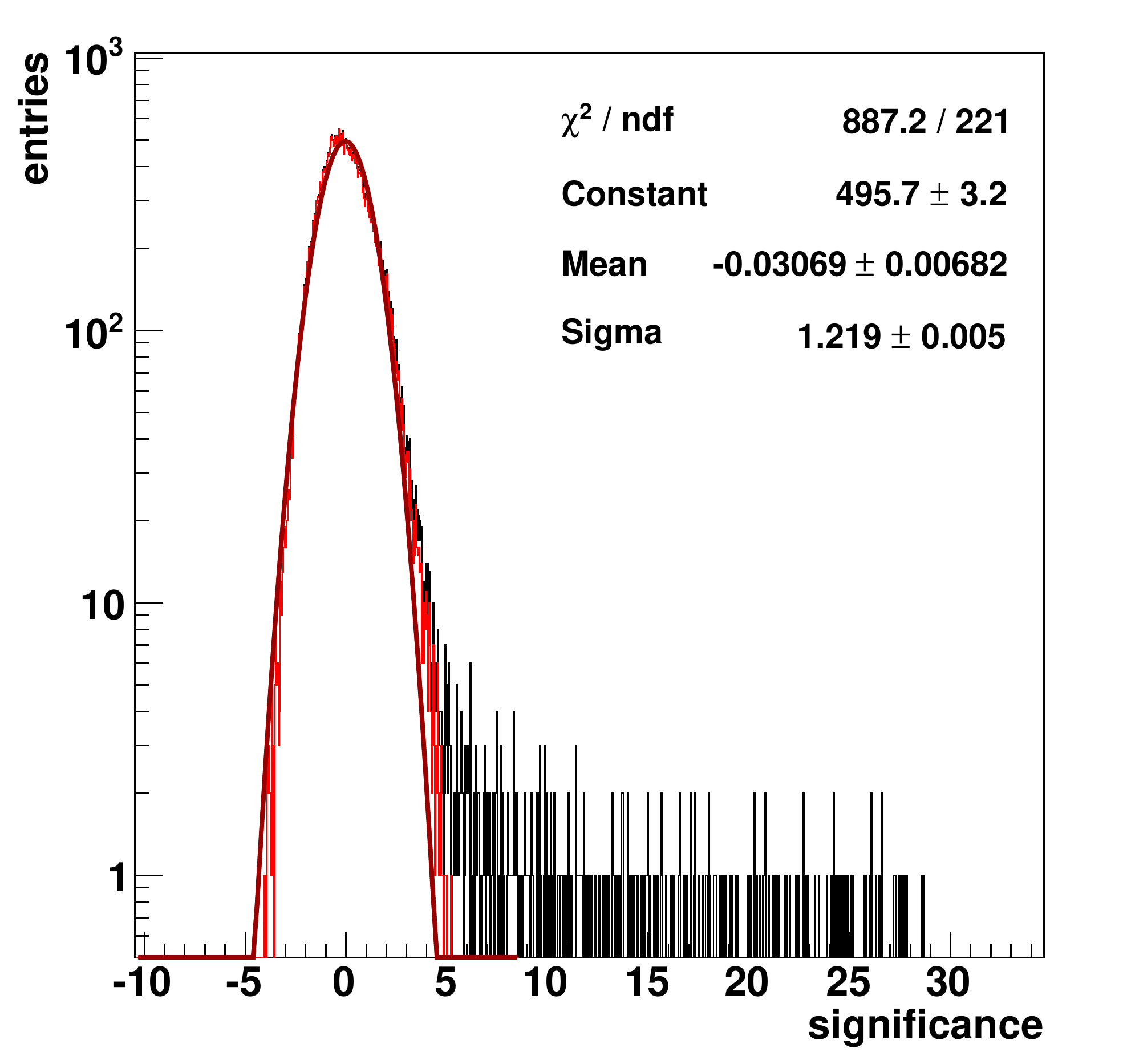}
  \caption{Left: A two dimensional distribution of excess events observed in the direction of PG 1553+113
using the CT5 Mono analysis (16.8 hr live time). The source position is indicated by magenta star.
Right: The significance distribution that corresponds to the excess map.
The meaning of the histograms and statistics data is the same as in Fig.~\protect\ref{fig:pks2155}.
}
  \label{fig:pg1553}
\end{figure}

\begin{figure}
  \centering
    \includegraphics[width=0.4\linewidth]{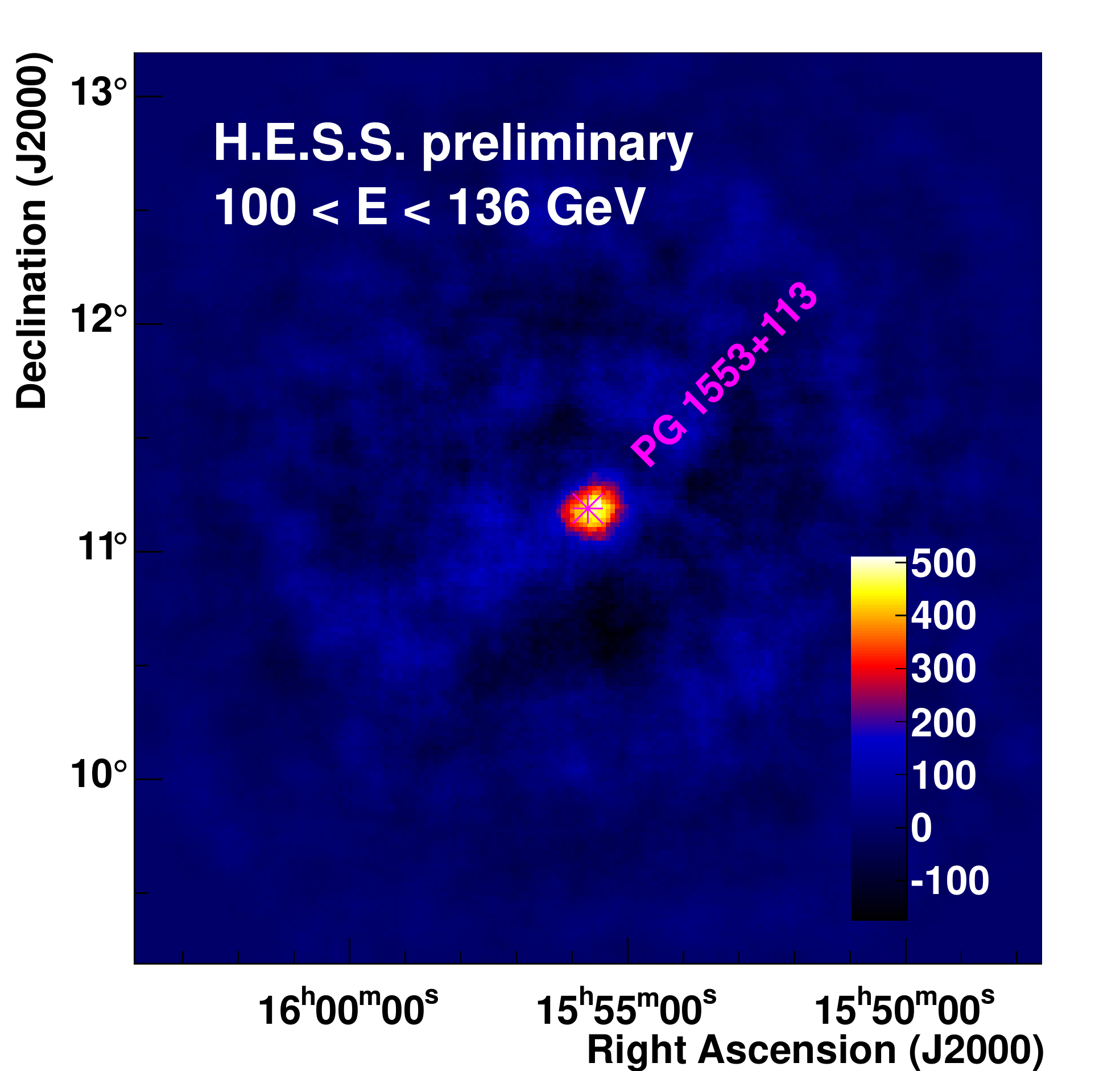}
    \includegraphics[width=0.4\linewidth]{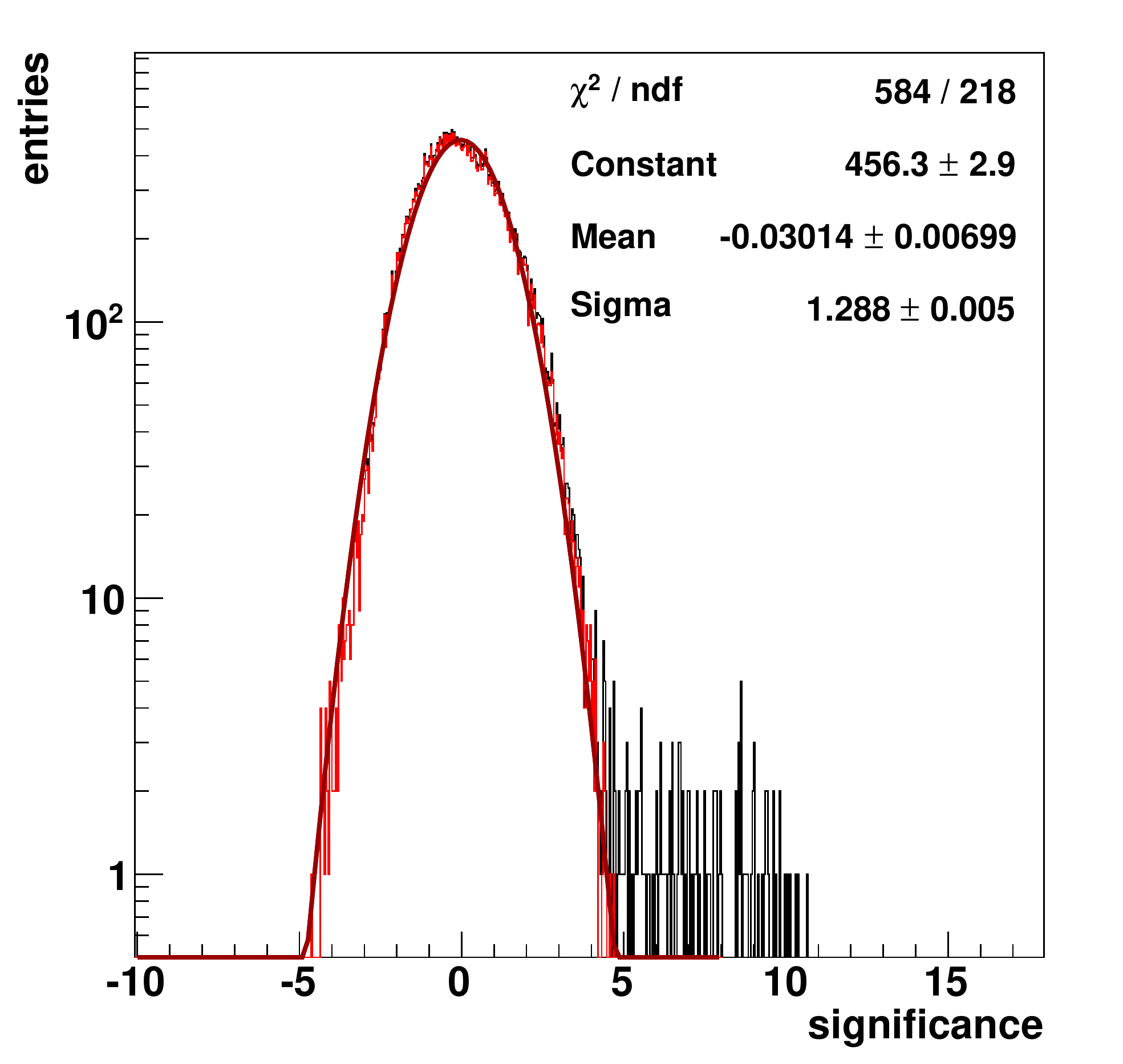}
  \caption{The PG 1553+113 excess map (left) and significance distribution (right) for events with reconstructed energy between 100 and 136~GeV
(CT5 Mono analysis).
}
  \label{fig:pg1553_ebins}
\end{figure}

\begin{figure}
  \centering
    \includegraphics[width=0.48\linewidth,height=0.31\linewidth]{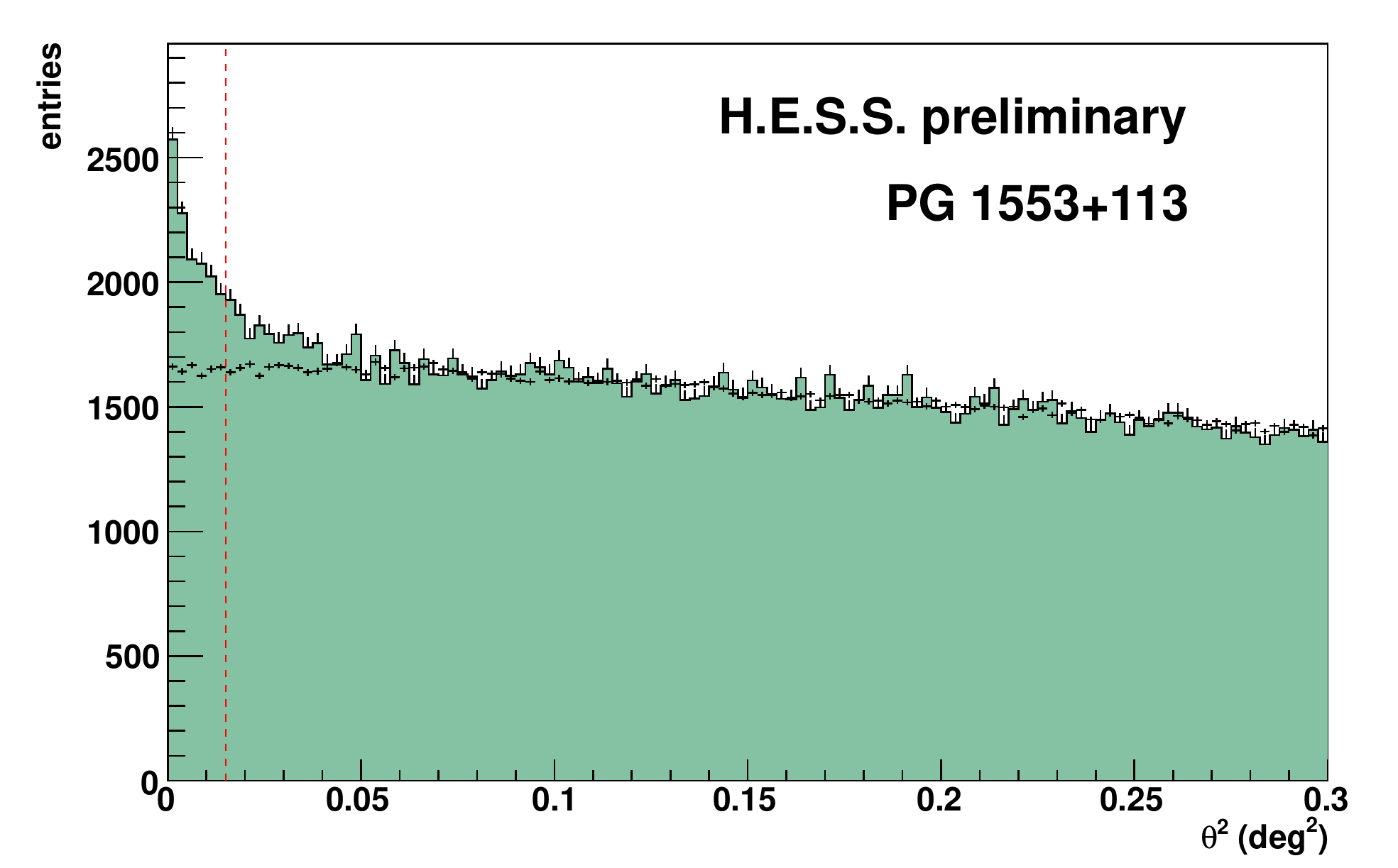}
    \includegraphics[width=0.5\linewidth]{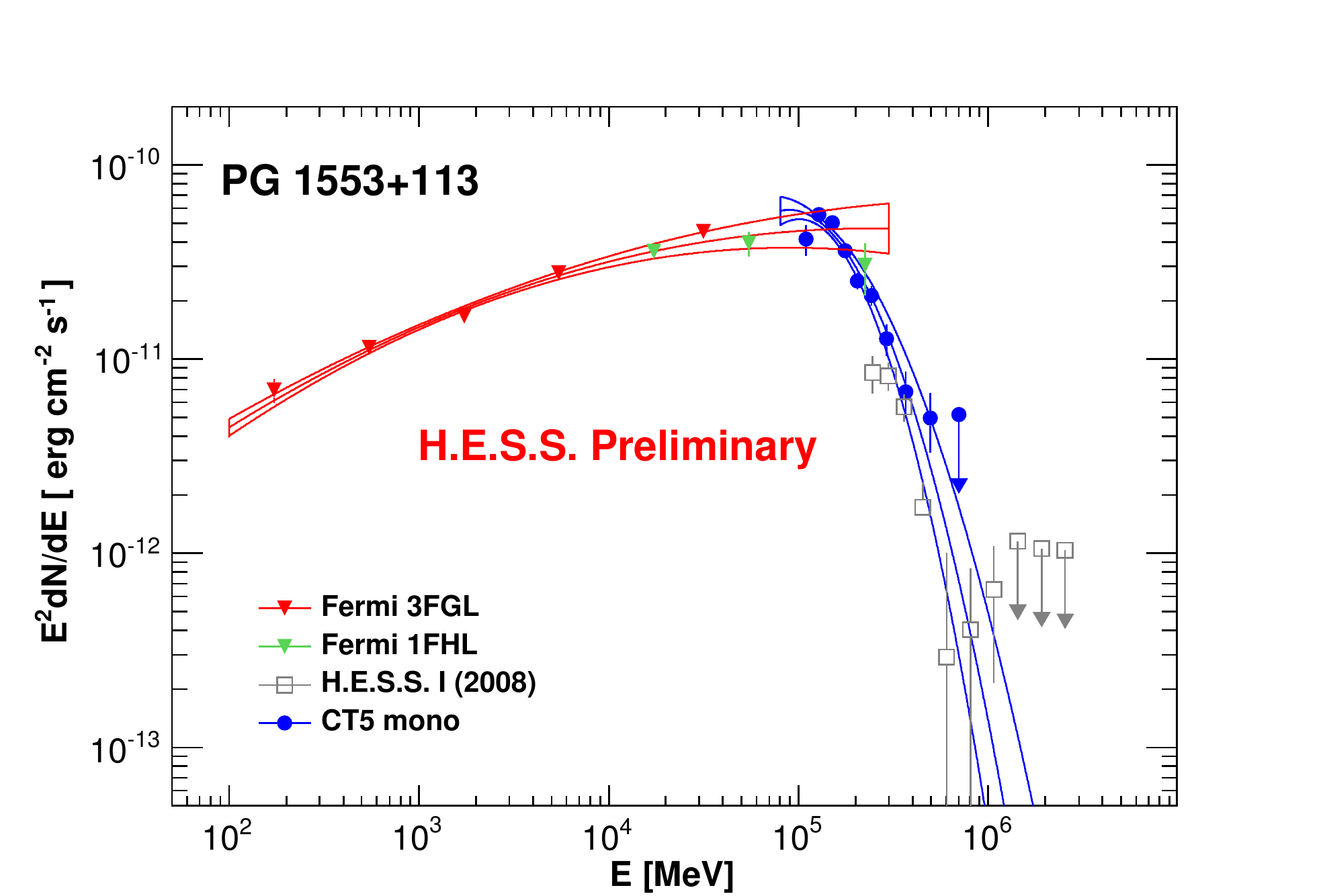}
  \caption{
Left: The $\theta^2$ distribution for PG 1553+113 (CT5 Mono analysis).
The meaning of the data shown is the same as in Fig.~\protect\ref{fig:pks2155_theta2}.
Right: The time-averaged energy spectrum of PG 1553+113 obtained from the CT5 Mono analysis
in comparison with the {\it Fermi} 3FGL and 1FHL catalog spectra \cite{3FGL,1FHL}
and the archival H.E.S.S. data from \protect\cite{hess_pg1553_spectroscopy}.
  }
  \label{fig:pg1553_theta2}
  \label{fig:pg1553_spectrum}
\end{figure}

The PG 1553+113 data set, filtered as explained in Sect.~\ref{sect:datasets}, comprises 39 data taking runs (16.8 hr live time),
which were analyzed using Loose cuts as described in Sect.~\ref{sect:analysis}.
The observation zenith angle was between $33^\circ$ to $40^\circ$, with a mean value of $35^\circ$.
The sky map obtained for PG 1553+113 using the CT5 Mono analysis is shown in Fig.~\ref{fig:pg1553}, left.
The source is detected at a high significance, with at least 2500 excess events.
The excess significance distribution outside of the $0.3^\circ$ exclusion radius is approximately consistent with a normal distribution (Fig.~\ref{fig:pks2155}, right).
The same holds true when the analysis is repeated in three energy bins equally spaced in logarithm of reconstructed energy between 100 and 250 GeV.
Using only events in the first energy bin ($100 < E_{\text{rec}} < 136$~GeV) 
the source is detected with a 10 $\sigma$ significance (Fig.~\ref{fig:pg1553_ebins}).
The significance distribution outside the exclusion region has $\sigma = 1.29$,
indicating that the background subtraction errors are smaller than the statistical errors.
The $\theta^2$ distribution is shown in Fig.~\ref{fig:pg1553_theta2}, left.
A 27.4 $\sigma$ excess over the background is observed within the on-source region ($\theta^2 < 0.015$ deg$^2$).
The reconstructed spectrum is well fitted by a log parabola (Fig.~\ref{fig:pg1553_spectrum}, right),
with a photon index $\Gamma = 2.95 \pm 0.23_{\text{stat}}$ 
at decorrelation energy $E_{dec} = 141$~GeV, curvature parameter $\beta = 1.04 \pm 0.31_{\text{stat}}$,
and differential flux $\frac{d\Phi}{dE} = (1.48 \pm 0.07_{\text{stat}}) \times 10^{-9} \, \text{cm}^{-2} \text{s}^{-1} \text{TeV}^{-1}$ at $E_{dec}$.
The systematic errors are currently under investigation.
The spectrum data points (blue filled circles) cover the energy range from 100~GeV to 500~GeV (not including upper limits).
The new measurement is in reasonable agreement with the earlier measurements by H.E.S.S.
\cite{hess_pg1553_spectroscopy,hess_pg1553_flare} (at E $>$ 200~GeV) and MAGIC \cite{MAGIC_pg1553_spectrum_curvature},
as well as with the {\it Fermi}-LAT catalog spectra (at E $<$ 200~GeV).

\section{Cross check analysis}\label{sect:crosscheck}
The robustness of the results presented above has been tested through an independent analysis
using the Image Pixel-wise fit for Atmospheric Cherenkov Telescopes (ImPACT) method described in \cite{ImPACT},
adapted for the reconstruction of data coming from CT5-only triggers.
The analysis was equally capable to detect PKS 2155--304 below 100 GeV
and the derived spectra were found in very good agreement with the Model analysis
for both PKS 2155--304 and PG 1553+113.

\section{Conclusion}\label{sect:conclusion}
An analysis of two bright AGN, PKS 2155--304 and PG 1553+113, using the H.E.S.S. CT5 data has been presented.
Both sources are detected with high significance, with an energy threshold of $\approx$ 80 and 100~GeV for PKS 2155--304 and PG 1553+113, respectively.
A spectral curvature has been confidently observed for both sources.
The measured spectrum of PKS 2155--304, using the 2013 and 2014 H.E.S.S. data, is approximately consistent with the quiescent spectrum reported earlier by H.E.S.S.,
as well as the {\it Fermi} 3FGL and 1FHL catalogs.
The obtained PG 1553+113 spectrum is in good agreement with earlier H.E.S.S. measurements and the {\it Fermi} catalogs.
All results presented in this paper are preliminary.

\end{document}